\documentclass[aps,pra,preprint,groupedaddress,showpacs]{revtex4}
\usepackage[dvips]{graphics}
\usepackage{graphicx}
\begin{document}
\title{Spin chains with electrons in Penning traps}
\author{G. Ciaramicoli, I. Marzoli, and P. Tombesi}
\affiliation{Dipartimento di Fisica, Universit\`a degli Studi di
Camerino, 62032 Camerino, Italy}
\date{\today}
\begin{abstract}
We demonstrate that spin chains are experimentally feasible using
electrons confined in micro-Penning traps, supplemented with local
magnetic field gradients. The resulting Heisenberg-like system is
characterized by coupling strengths showing a dipolar decay. These
spin chains can be used as a channel for short distance quantum
communication. Our scheme offers high accuracy in reproducing an
effective spin chain with relatively large transmission rate.
\end{abstract}
\maketitle
\section{Introduction}
Recently much theoretical research work has been focused on the possibility to
use systems of spins, coupled by ferromagnetic Heisenberg
interactions and arranged along chain structures, for transferring
quantum information. The remarkable property of these systems is
the capability of transmitting the qubit state along the chain
with fidelity exceeding the classical threshold and
by means only of their free dynamical evolution. After the seminal
paper by Bose \cite{bose}, in which the potentialities of the
so-called spin chains have been shown, several strategies have been proposed to
increase the transmission fidelity \cite{osborne} and
even to achieve, under appropriate conditions, perfect state transfer
\cite{chri1,chri2,burgarth1,burgarth2}. All these proposals
refer to ideal spin chains in which only nearest-neighbor couplings are
present. However, also the more realistic case of long range
couplings, in particular magnetic dipole like couplings, has been
studied \cite{kay,avellino}. In \cite{kay} it has been shown that perfect
state transfer or, at least, high transmission fidelity can be
obtained by appropriately choosing the system parameters, such as
local magnetic fields and inter-spin distances.
Moreover, even when no site specific
locally-tunable fields are allowed, spin chains with
dipolar couplings often perform better, in terms of
transmission fidelity, than their
nearest-neighbor coupled counterpart \cite{avellino}.
Hence, from these theoretical predictions,
we expect that spin chains, also in the case of long
range interactions, may represent a very promising system to achieve
high fidelity quantum information transfer without requiring
experimentally demanding gating operations.

In this paper we demonstrate that a linear array of electrons,
confined in micro-Penning traps, can implement an effective spin
chain with magnetic dipole like spin coupling. The same system
consisting of trapped electrons in vacuum has been already
proposed as a valid and competitive candidate for universal
quantum information processing
\cite{ciaramicoli4,ciaramicoli_5,nuovo}. In this respect, the
possibility of reliably transmitting the qubit state between
different quantum registers, without applying gate operations, is
highly desirable.
In fact, the use of a quantum channel to transfer a qubit state
in a quantum processor can be a valuable alternative to the
repeated application of swapping gates.

We have already proved in \cite{nuovo}
that the addition of a magnetic field gradient, together with the
Coulomb interaction between the particles, allows to obtain an
effective nuclear magnetic resonance (NMR)-like coupling between
the spins of the confined electrons. Here we generalize this
approach to encompass a variety of trap set up, also in connection
with novel geometries of Penning traps \cite{wiretrap,stahl}.
Indeed, by further investigating the
interaction between the internal (spin) and external (motional)
degrees of freedom of each particle, introduced by the applied local
magnetic field gradient, we can mimic more general systems, with
Heisenberg ferromagnetic or antiferromagnetic Hamiltonian. This
fact opens up the possibility to simulate quantum spin systems with
tunable interactions, thanks to the experimental control over the
different trap parameters. The ultimate goal may be the
observation of quantum phase transitions, as proposed with trapped
ions controlled by laser beams \cite{porras}.

In our proposal, we consider a linear array of electrons with
inter-particle distance ranging from few microns to 50~$\mu$m.
We provide an analytical expression for the spin-spin coupling
strength, which shows a dipolar decay law.
We estimate the value of the spin-spin
coupling, for different ranges of the system characteristic frequencies
as well as of the intensity of the magnetic gradient, with the aim
of optimizing the transfer time of our quantum channel.
Furthermore, we evaluate the fidelity of our system in reproducing
an effective spin chain according to the Heisenberg model.
In particular, we calculate the
probability to obtain a perfect spin state transfer
in a chain consisting of just two electrons. This
probability, equal to one in an ideal spin chain \cite{bose}, in
our system is less than one owing to the effects resulting from
the interplay between the internal and the external degrees of freedom
of the trapped particles.
However, by an appropriate choice of the
system parameters, especially the frequency and the amplitude of
the spatial motions, we can obtain high fidelities and, at the
same time, sufficiently large values of the spin-spin coupling.
The electron trapping arrangement offers also the possibility to
apply arbitrary site-specific changes in the system parameters in
order to maximize, as outlined in \cite{kay,avellino}, the
efficiency of the quantum channel. Our theoretical predictions suggest
that a linear array of electrons is suitable to implement a spin
chain with the present technology.

The manuscript is organized as follows.
In Sec.~\ref{system} we describe the system and
how the local magnetic field gradient
couples the electron spin to the motional degrees of freedom.
This coupling, mediated by the Coulomb interaction between charged
particles, results in an effective Heisenberg-like Hamiltonian
(Sec.~\ref{spin}).
In Sec.~\ref{chain} we estimate the fidelity and the efficiency
of our system as a channel for quantum information transmission.
Finally, the results of our analysis
are summarized and discussed in Sec.~\ref{concl}.
The more technical details, concerning the derivation of the fidelity,
are presented in Appendix~\ref{appendix}.
\section{A linear array of trapped electrons}
\label{system}
\begin{figure}
 \includegraphics{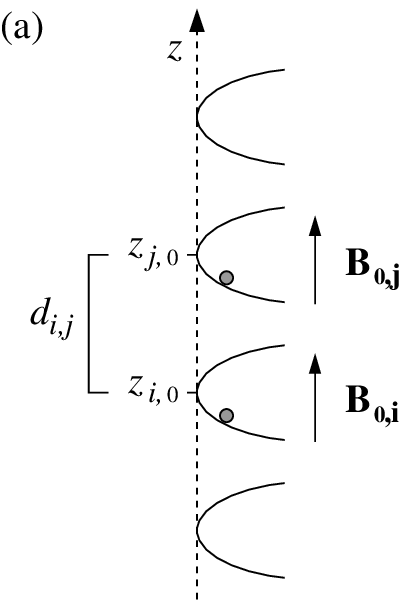}%
 \includegraphics{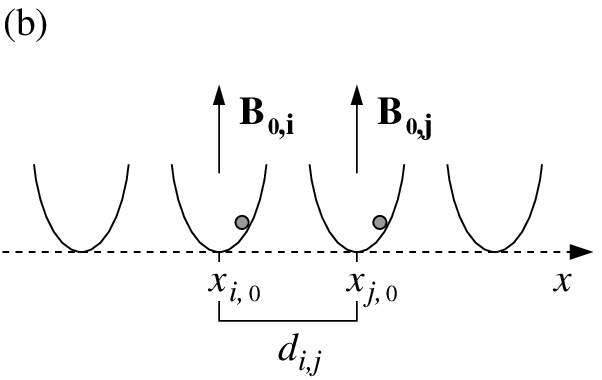}
 \caption{\label{array} Schematic drawing illustrating
  two different geometries for
 a linear array of micro-Penning traps. The traps are represented
 by sketching the electrostatic potential along the $z$ axis.
 (a) The electrons are aligned
 along the $z$ axis, parallel to  the confining magnetic
field; (b) the electrons are aligned along the $x$ axis,
orthogonal to the confining magnetic field.}
\end{figure}

We consider a linear array of $N$ electrons in micro-Penning
traps \cite{drndic}.
According to the different geometry of the electrode
arrangement, the micro-trap array can be either parallel to the
direction of the confining magnetic field, i.e. along the $z$ axis
as shown in Fig.~\ref{array}(a), or orthogonal to
this field, for example, along the $x$ axis as shown in Fig.~\ref{array}(b).
To confine electrons in an array along the $z$ direction we can use
a closed cylindrical electrode structure
\cite{ciaramicoli4,ciaramicoli_5} or an open wire arrangement
\cite{wiretrap}.
This latter structure can also accommodate the
electrons in an array aligned along the $x$ axis.
An orderly set of
micro-traps, orthogonal to the trapping magnetic
field, can be likewise realized by means of a planar electrode system
\cite{stahl}.
As we will see, the different orientation of the
linear array of particle affects the form of the resulting interaction
Hamiltonian.
Hence, we firstly derive the
expression of the effective Hamiltonian in the case of
micro-traps aligned along the $z$ axis. Then we will show how this
expression modifies in the case of an array directed along the
$x$ axis.

The Hamiltonian of a system of $N$ electrons confined in an array of
Penning traps can be written as
\begin{equation}  \label{Harray}
   H = \sum_{i=1}^{N} H_i^{NC} + \sum_{i < j} H_{i,j}^{C},
\end{equation}
where
\begin{equation} \label{H_NC}
   H_i^{NC} = \frac{(\mathbf{p_i} -e \mathbf{A_i})^2}{2m_e} + eV_i
                    - \frac{ge\hbar}{4m_e} \mbox{\boldmath $\sigma_i$} \cdot
                    \mathbf{B_i}
\end{equation}
represents the single electron dynamics inside each trap and
\begin{equation} \label{H_C}
    H_{i,j}^{C} = \frac{e^2}{4 \pi \epsilon_0|\mathbf{r_i}-\mathbf{r_j}|}
\end{equation}
describes the Coulomb interaction between electrons $i$ and $j$.
In Eqs.~(\ref{H_NC}) and (\ref{H_C}) $m_e$, $e$, $g$, and {\boldmath $\sigma_i$}
are, respectively, the electron mass,
charge, giromagnetic factor, and Pauli spin operators.
As shown in Fig.~\ref{array}(a), we assume that
the micro-traps are aligned along the $z$ axis and that $z_{i,0}$
is the position of the center of the $i$-th trap.
The electrostatic potential
\begin{equation}
    V_i(x_i, y_i, z_i) \equiv V_0 \,
                              \frac{(z_i-z_{i,0})^2-(x_i^2+y_i^2)/2}{\ell^2}
\end{equation}
is the usual quadrupole potential of a Penning trap,
where
$V_0$ is the applied potential difference between the trap electrodes
and $\ell$ is a characteristic trap length.
The magnetic field
\begin{equation} \label{magfield2}
\mathbf{B_i} \equiv - \frac{b}{2} (x_i \, \mathbf{i} + y_i \, \mathbf{j})
                 + \left[ B_0 + b (z_i-z_{i,0}) \right] \mathbf{k}
\end{equation}
is the sum of the uniform magnetic field $B_0 \, \mathbf{k}$,
providing the radial confinement, with a local linear magnetic
gradient around the $i$-th trap. The associated vector potential
\begin{equation}
   \mathbf{A_i} \equiv \frac{1}{2}
          [B_{0} + b (z_i-z_{i,0})] ( - y_i \, \mathbf{i} + x_i \, \mathbf{j} )
\end{equation}
preserves the cylindrical symmetry of the unperturbed trapping field.

Following an approach similar to the one described in Ref.~\cite{nuovo}, the
spatial part of $H_i^{NC}$ can be recast in the form
\begin{eqnarray} \label{Hspa}
  H_i^{(ext)} & \simeq & -\hbar\omega_{m} a_{m,i}^\dagger a_{m,i}
    + \hbar\omega_{c} a_{c,i}^\dagger a_{c,i}
    + \hbar\omega_z a_{z,i}^\dagger a_{z,i} \nonumber \\
    &+& \hbar \omega_z
      \varepsilon
       (a_{z,i}+a_{z,i}^{\dagger})\left(\frac{\omega_{m}}{\omega_{c}}
        a_{m,i}^\dagger a_{m,i}
        + a_{c,i}^\dagger a_{c,i} \right),
\end{eqnarray}
where the annihilation operators $a_{m,i}$, $a_{c,i}$, $a_{z,i}$
\cite{nuovo,brown} refer, respectively, to the magnetron,
cyclotron and axial oscillators of the $i$-th electron. The
frequencies of these oscillators
\begin{eqnarray}
   \omega_{m} &\simeq & \frac{\omega_z^2}{2 \omega_{c}}, \\
   \omega_{c} &\simeq & \frac{|e|B_{0}}{m_e}, \\
   \omega_z &=& \sqrt{ \frac{2e V_0}{m_e \ell^2}}
\end{eqnarray}
depend on the applied external fields and on the trap size.
They build up a well defined hierarchy with
$\omega_{m} \ll \omega_z \ll \omega_{c}$.
Indeed, we exploit this fact together with the assumption of a weak
magnetic gradient, such that $b|z_i - z_{i,0}|/B_0 \ll 1$, to derive
the Hamiltonian (\ref{Hspa}).
The dimensionless parameter
\begin{equation}
   \varepsilon \equiv \frac{|e|b}{m_e \omega_z}
               \sqrt{ \frac{\hbar}{2 m_e \omega_z} }
               = \frac{|e|b \Delta z}{m_e \omega_z},
\end{equation}
with $\Delta z$ being the ground state amplitude of the axial
oscillator, represents the coupling, due to the magnetic gradient,
between the axial motion and the radial degrees of freedom. In a
similar way the magnetic gradient introduces also an interaction
between the spatial and the spin motion. This coupling becomes evident
by considering the spin part of the Hamiltonian Eq.~(\ref{H_NC})
\begin{eqnarray}\label{Hs1}
  H_i^{(spin)} &\equiv & -\frac{ge\hbar}{4m_e}
                         \mbox{\boldmath $\sigma_i$ } \cdot \mathbf{B_i}
  \nonumber \\
               & = & \frac{g \hbar}{4}\omega_{c} \sigma^z_i +
   \frac{g\hbar|e|b}{4 m_e} \sigma^z_i (z_i-z_{i,0})
  -\frac{g\hbar|e|b}{8 m_e}(\sigma^x_{i} x_i+\sigma^y_{i} y_i)
\end{eqnarray}
which, in terms of the ladder operators, becomes \cite{nuovo}
\begin{equation}\label{Hs2}
  H_i^{(spin)} \simeq \frac{\hbar}{2}\omega_{s} \sigma^z_{i}
                      + \frac{g}{4} \varepsilon \hbar \omega_{z}
                        \sigma^z_{i} (a_{z,i}+a_{z,i}^\dagger)
  - \frac{g}{4} \varepsilon \hbar \omega_{z}
    \sqrt{\frac{\omega_{z}}{\tilde{\omega}_{c}}}
  \left[ \sigma^{(+)}_{i} (a_{c,i}+a_{m,i}^\dagger)
   +\sigma^{(-)}_{i} (a_{c,i}^\dagger+a_{m,i}) \right],
\end{equation}
where $\tilde{\omega}_c \equiv \sqrt{\omega_c^2 - 2 \omega_z^2}$ is
essentially a modified cyclotron frequency due to the insertion of
the quadrupole potential. In deriving Eq. (\ref{Hs2}) we defined
the operators $\sigma^{(\pm)}_{i}\equiv (\sigma^x_i \pm i
\sigma^y_i)/2$ and the spin precession frequency
$\omega_{s}\equiv(g/2)\omega_{c}$.

Hence, the Hamiltonian, Eq.~(\ref{H_NC}), of a single electron can be written as
\begin{eqnarray}  \label{Hglo3}
H_i^{NC} &\simeq& -\hbar\omega_{m} a_{m,i}^\dagger a_{m,i}
    + \hbar\omega_{c} a_{c,i}^\dagger a_{c,i}
    + \hbar\omega_{z} a_{z,i}^\dagger a_{z,i}+\frac{\hbar}{2}\omega_{s}
    \sigma^z_{i}
    \nonumber \\
   &+&\hbar \omega_z \varepsilon (a_{z,i}+a_{z,i}^{\dagger})
       \left(a_{c,i}^\dagger
   a_{c,i} + \frac{\omega_{m}}{\omega_{c}}a_{m,i}^\dagger
   a_{m,i}+\frac{g}{4}
\sigma^z_{i}\right)\nonumber \\
   &-& \frac{g}{4} \varepsilon \hbar \omega_z
       \sqrt{\frac{\omega_z}{\tilde{\omega}_{c}}}
    \left[ \sigma^{(+)}_{i} (a_{c,i}+a_{m,i}^\dagger)
          +\sigma^{(-)}_{i} (a_{c,i}^\dagger+a_{m,i})
    \right] .
\end{eqnarray}
We now assume that, for each particle of the array, the cyclotron
oscillator is in the ground state and the magnetron oscillator is
in a thermal state with an average excitation number $\bar{l} \ll
\omega_c / \omega_m$. We recall that the ground state cooling of the
cyclotron motion for electrons \cite{gabrielse} and the reduction of
the magnetron motion excitation number for electrons \cite{brown}
and ions \cite{powell} have been experimentally obtained. Under
these conditions, we can neglect in Eq.~(\ref{Hglo3}) the coupling
between the axial oscillator and the radial
motion. We can further simplify Eq.~(\ref{Hglo3}) by means of the
rotating wave approximation (RWA). Indeed, terms like
$\sigma^{(+)}_{i} a_{m,i}^\dagger$ are rotating at a frequency
$\omega_s - \omega_m$ much larger than the anomaly frequency
$\omega_a \equiv \omega_s - \omega_c$
typical of terms like $\sigma^{(+)}_{i} a_{c,i}$
and, therefore, are negligible in RWA. Hence, the Hamiltonian
Eq.~(\ref{Hglo3}) becomes
\begin{eqnarray} \label{H_NC_rwa}
H_i^{NC} &\simeq& -\hbar\omega_{m} a_{m,i}^\dagger a_{m,i}
    + \hbar\omega_{c} a_{c,i}^\dagger a_{c,i}
    + \hbar\omega_{z} a_{z,i}^\dagger a_{z,i}+\frac{\hbar}{2}\omega_{s}
    \sigma^z_{i}
    \nonumber \\
   &+& \frac{g}{4} \varepsilon \hbar \omega_z
       \left( a_{z,i}+a_{z,i}^{\dagger} \right) \sigma^z_{i}
        - \frac{g}{4} \varepsilon \hbar \omega_z
       \sqrt{\frac{\omega_z}{\tilde{\omega}_{c}}}
       \left( \sigma^{(+)}_{i} a_{c,i} + \sigma^{(-)}_{i}  a_{c,i}^\dagger
       \right).
\end{eqnarray}
We see that the applied magnetic field gradient couples the different
electron spin components to the axial
and to the cyclotron oscillators.

Let us consider the part of the Hamiltonian Eq.~(\ref{H_C}) describing
the Coulomb interaction between two electrons trapped at the sites $i$
and $j$.
If the oscillation amplitude of the
two electrons is much smaller than the inter-trap distance
$d_{i,j}\equiv|z_{i,0}-z_{j,0}|$, we can expand the
interaction Hamiltonian
in a power series and retain the
terms up to the second order
\begin{equation} \label{HC3}
  H^{C}_{i,j} \simeq - \frac{e^2}{4 \pi \epsilon_0 d_{i,j}^2}
                     (\tilde{z}_i - \tilde{z}_j) +
  \frac{e^2}{8 \pi \epsilon_0 d_{i,j}^3}
   \left[ 2 (\tilde{z}_i - \tilde{z}_j)^2-(x_i-x_j)^2 -
              (y_i-y_j)^2\right] ,
\end{equation}
where $\tilde{z}_i \equiv z_i-z_{i,0}$.
The Coulomb interaction produces three effects on the electron dynamics:
(i) a displacement of the equilibrium position along the $z$ axis,
(ii) a shift of the axial resonance frequency, and
(iii) a coupling between the motional degrees of freedom of different particles.
The first two effects are rather small and can be taken into account by
redefining the trap center position and the corresponding axial
frequency.
Therefore, in the remainder of this section we focus on the coupled dynamics
of the two electrons
\begin{eqnarray} \label{HC4}
  H^{C}_{i,j} &\simeq& -\frac{e^2}{4 \pi \epsilon_0 d_{i,j}^3} (2 \tilde{z}_i
  \tilde{z}_j -x_i x_j -y_i y_j) \nonumber\\
&=& -2 \hbar \xi_{i,j} (a_{z,i}+a^{\dagger}_{z,i})(a_{z,j}+a^{\dagger}_{z,j})
     \nonumber \\
&+& \hbar \xi_{i,j} \frac{\omega_z}{\tilde{\omega}_{c}}
       (a_{c,i}+a_{c,i}^\dagger+a_{m,i}+a_{m,i}^\dagger)
       (a_{c,j}+a_{c,j}^\dagger+a_{m,j}+a_{m,j}^\dagger) \nonumber \\
&-& \hbar \xi_{i,j} \frac{\omega_z}{\tilde{\omega}_{c}}
       (a_{c,i}-a_{c,i}^\dagger-a_{m,i}+a_{m,i}^\dagger)
       (a_{c,j}-a_{c,j}^\dagger-a_{m,j}+a_{m,j}^\dagger) ,
\end{eqnarray}
where the coupling strength
\begin{equation}
\xi_{i,j} \equiv \frac{e^2}{8 \pi \epsilon_0 m_e \omega_z
         d_{i,j}^3}
          = \frac{1}{\hbar} \frac{e^2}{4  \pi \epsilon_0 d_{i,j}}
            \left( \frac{\Delta z}{d_{i,j}} \right)^2
\end{equation}
amounts to the Coulomb energy times the square of the ratio
between the axial ground state amplitude and the inter-particle
distance. We have observed that each oscillator, axial, cyclotron
and magnetron, is characterized by a typical resonance frequency.
As a consequence, the coupling introduced by the Coulomb
interaction between the degrees of freedom of different electrons
is effective only for almost resonant oscillators. Therefore, in
Eq.~(\ref{HC4}) the terms that couple the cyclotron and magnetron
motion of the two particles give negligible effects. Furthermore,
we are not interested in the coupling between the magnetron motion
of different electrons, since this mode is essentially
decoupled from the other degrees of freedom. Hence, disregarding
the magnetron motion, the part of the system Hamiltonian
describing the Coulomb repulsion between electrons $i$ and $j$
reduces to
\begin{eqnarray}  \label{HC5}
  H^{C}_{i,j} & \simeq & -2 \hbar \xi_{i,j}
  (a_{z,i}+a^{\dagger}_{z,i})(a_{z,j}+a^{\dagger}_{z,j})
+ 2 \hbar \xi_{i,j}
\frac{\omega_z}{\tilde{\omega}_c} \left( a_{c,i} a_{c,j}^\dagger
                                +a_{c,i}^\dagger a_{c,j}
                          \right).
\end{eqnarray}
In the case of a linear array of electrons, trapped in a direction
orthogonal to the magnetic field, i.e. along the $x$ axis as shown in
Fig.~\ref{array}(b), we can derive a similar
expression for the Coulomb coupling
\begin{eqnarray}\label{HC6}
  H^{C}_{i,j} & \simeq & \hbar \xi_{i,j}
  (a_{z,i}+a^{\dagger}_{z,i})(a_{z,j}+a^{\dagger}_{z,j})
- \hbar \xi_{i,j} \frac{\omega_z}{\tilde{\omega}_c}
   \left[ a_{c,i} a_{c,j}^\dagger + a_{c,i}^\dagger a_{c,j} \right.
\nonumber \\
 & + & \left. 3 ( a_{c,i} a_{c,j} + a_{c,i}^\dagger a_{c,j}^\dagger ) \right].
\end{eqnarray}
We emphasize that in the case of Eq.~(\ref{HC5}), referring to the
vertical array of traps, the coupling between the cyclotron
oscillators of different electrons represent a swapping of
excitations, which basically conserves energy. The only terms that
survive involve the creation of a quantum of excitation at the
site $j$ at the expense of the destruction of a quantum of
excitation at the site $i$ and viceversa. In the case of an
horizontal arrangement of traps, Eq.~(\ref{HC6}), this is, in
general, no longer true. Indeed, even though the leading terms are
preserving the energy of the two coupled cyclotron oscillators, we
also note the presence of rapidly rotating terms which involve the
simultaneous creation and annihilation of two excitations. However
if $\xi_{i,j} (\omega_z/\tilde{\omega}_c) \ll \omega_c$ the effects of
these rapidly rotating terms are negligible (RWA)
and the Hamiltonian (\ref{HC6}) becomes
\begin{eqnarray}\label{HC7}
  H^{C}_{i,j} & \simeq & \hbar \xi_{i,j}
  (a_{z,i}+a^{\dagger}_{z,i})(a_{z,j}+a^{\dagger}_{z,j})
- \hbar \xi_{i,j} \frac{\omega_z}{\tilde{\omega}_c}
   \left( a_{c,i} a_{c,j}^\dagger + a_{c,i}^\dagger a_{c,j} \right).
\end{eqnarray}
We also note that Eqs.~(\ref{HC5}) and (\ref{HC7}) exhibit alternating signs
in front of the coupling terms.
As we will see in the next section, this
results in a different kind, ferromagnetic or antiferromagnetic, of
the effective spin-spin interaction.
\section{Effective spin-spin coupling} \label{spin}
In the previous section we have seen that the magnetic gradient
induces, for each particle of the array, a coupling between the
spatial and the spin motions. This interaction, mediated by the
Coulomb repulsion between the electrons, gives rise to an
effective spin-spin coupling between different particles \cite{nuovo}. This
effect can be brought to light by making an appropriate unitary
transformation on the Hamiltonian of the system \cite{wunderlich}.
We seek a
transformation that formally removes, in the single particle
Hamiltonian, the coupling
between the internal and the external degrees of freedom of each electron.
Hence, we transform the Hamiltonian,
Eq.~(\ref{Harray}), as $H'=e^S H e^{-S}$ with
\begin{eqnarray} \label{S}
S &=& \frac{g}{4} \, \varepsilon \sum_{i=1}^N
  \left[\sigma^z_{i} (a_{z,i}^\dagger-a_{z,i})
        +\frac{\omega_z}{\omega_{a}}
         \sqrt{\frac{\omega_z}{\tilde{\omega}_{c}}}
        \left( \sigma^{(-)}_{i} a_{c,i}^\dagger -\sigma^{(+)}_{i} a_{c,i}
        \right)
  \right] ,
\end{eqnarray}
where $\omega_{a}\equiv\omega_{s}-\omega_{c}$ is the anomaly frequency.
The unitary transformation changes the operators, to the
first order in $\varepsilon$,
in the following way
\begin{eqnarray}
a_{z,i} &\rightarrow& a_{z,i}-\frac{g}{4} \varepsilon \, \sigma^z_i,\\
a_{c,i} &\rightarrow& a_{c,i}-\frac{g}{4} \varepsilon
                     \frac{\omega_z}{\omega_{a}}
            \sqrt{\frac{\omega_z}{\tilde{\omega}_{c}}} \, \sigma^{(-)}_{i}, \\
\sigma^z_i &\rightarrow& \sigma^z_i+\frac{g}{2} \varepsilon
        \frac{\omega_z}{\omega_{a}} \sqrt{\frac{\omega_z}{\tilde{\omega}_{c}}}
  \left( \sigma^{(+)}_{i} a_{c,i}+\sigma^{(-)}_{i} a_{c,i}^\dagger \right), \\
\sigma^{(+)}_{i} &\rightarrow&  \sigma^{(+)}_{i}+ \frac{g}{2}
\varepsilon \, \sigma^{(+)}_{i} (a_{z,i}^\dagger-a_{z,i})-\frac{g}{4}
\varepsilon
        \frac{\omega_z}{\omega_{a}} \sqrt{\frac{\omega_z}{\tilde{\omega}_{c}}}
        \, \sigma^z_{i} a_{c,i}^\dagger \, .
\end{eqnarray}
To derive the expressions above we made use of the expansion
\begin{equation}
e^{\eta A} B e^{-\eta A}=B+\eta[A,B]+\frac{\eta^2}{2!} [A,[A,B]]+
\frac{\eta^3}{3!} [A,[A,[A,B]]]+ \ldots ,
\end{equation}
where $A$ and $B$ are two noncommuting operators and $\eta$ is a
parameter. \\
The single electron part, Eq.~(\ref{H_NC_rwa}), of the system Hamiltonian
can be written, after applying the
unitary transformation, as
\begin{eqnarray}\label{HNCp}
H_{i}^{'NC} \simeq  -\hbar\omega_{m} a_{m,i}^\dagger a_{m,i}
    + \hbar\omega_{c} a_{c,i}^\dagger a_{c,i}
+ \hbar\omega_z a_{z,i}^\dagger a_{z,i}+ \frac{\hbar}{2}\omega_{s}
\sigma^z_{i} ,
\end{eqnarray}
where we have neglected second and higher order terms in
$\varepsilon$, which in the cases relevant to the present analysis
is of the order of 10$^{-2}$.
Nevertheless, these extra terms are derived in Appendix~\ref{appendix} and their
influence on the performances of the system is discussed in
Sec.~\ref{chain}.

Let us now turn to the Coulomb part of the system Hamiltonian.
The first term in
Eq.~(\ref{HC5}) becomes
\begin{eqnarray} \label{Hzij}
-2 \hbar \xi_{i,j}
  \left( a_{z,i}+a^{\dagger}_{z,i}-\frac{g}{2} \varepsilon \sigma^z_i
  \right)
  \left( a_{z,j}+a^{\dagger}_{z,j}-\frac{g}{2} \varepsilon \sigma^z_j
  \right) .
\end{eqnarray}
Expression (\ref{Hzij}) contains a term which represents an effective
spin-spin coupling between different electrons in the array.
This effect was already pointed out in Ref.~\cite{nuovo}.
Moreover, we note that the unitary transformation enforces a coupling
between the axial motion of the $j$-th electron and the spin of the
$i$-th electron.
This effect is smaller of a factor $\xi_{i,j} / \omega_z \ll 1$
than the corresponding coupling
[see Eq.~(\ref{H_NC_rwa})]
between internal (spin) and external (axial motion) degrees of freedom of the
same particle.
The error introduced by neglecting these terms is estimated in the Appendix.

The remaining term in Hamiltonian (\ref{HC5})
transforms into
\begin{eqnarray}\label{Hxij}
& & 2 \hbar \xi_{i,j} \frac{\omega_z}{\tilde{\omega}_{c}}
    \left( a_{c,i}-\frac{g}{4} \varepsilon \frac{\omega_z}{\omega_{a}}
           \sqrt{\frac{\omega_z}{\tilde{\omega}_{c}}}\sigma^{(-)}_{i}
    \right)
    \left( a_{c,j}^\dagger-\frac{g}{4} \varepsilon \frac{\omega_z}{\omega_{a}}
           \sqrt{\frac{\omega_z}{\tilde{\omega}_{c}}}\sigma^{(+)}_{j}
    \right)
\nonumber \\
&+& 2 \hbar \xi_{i,j} \frac{\omega_z}{\tilde{\omega}_{c}}
    \left( a_{c,i}^\dagger-\frac{g}{4} \varepsilon \frac{\omega_z}{\omega_{a}}
           \sqrt{\frac{\omega_z}{\tilde{\omega}_{c}}}\sigma^{(+)}_{i}
    \right)
    \left( a_{c,j}-\frac{g}{4} \varepsilon \frac{\omega_z}{\omega_{a}}
           \sqrt{\frac{\omega_z}{\tilde{\omega}_{c}}}\sigma^{(-)}_j
    \right) .
\end{eqnarray}
From Eq.~(\ref{Hxij}) we see that the unitary
transformation  produces the term
\begin{equation} \label{xycoup}
\hbar \xi_{i,j} \varepsilon^2 \frac{g^2}{8}
\frac{\omega_z^4}{\omega_a^2 \tilde{\omega}_c^2}
\left( \sigma^{(-)}_i \sigma^{(+)}_j + \sigma^{(+)}_i \sigma^{(-)}_j
\right) = \hbar \xi_{i,j}
\varepsilon^2 \frac{g^2}{16} \frac{\omega_z^4}{\omega_a^2
\tilde{\omega}_c^2}(\sigma^x_i \sigma^x_j + \sigma^y_i \sigma^y_j) ,
\end{equation}
which represents a direct coupling between the spin
motion of different particles.
Also in this case, there are additional terms in expression (\ref{Hxij}),
that couple the spin of an electron to the cyclotron motion of the
other electrons in the chain.
In comparison with the spin-cyclotron interaction for the same particle
[see Eq.~(\ref{H_NC_rwa})], this coupling is reduced of
a factor $\xi_{i,j} \omega_z / \tilde{\omega}_c \omega_a$, which, in the
range of the parameters considered here, is typically much less than one.
For an estimate of the errors introduced by these terms we refer to the
Appendix.

Hence, summarizing the results of our derivation, we have an
effective spin-spin coupling between the electrons with the spatial
dynamics substantially decoupled from the spin dynamics.
Consequently the spin part of the system Hamiltonian can be
written, in the case of a linear array of electrons along
the $z$ axis, as
\begin{eqnarray}\label{Heff1}
H'_s &\simeq& \sum_{i=1}^{N}\frac{\hbar}{2}\omega_{s} \sigma^z_i
- \hbar \sum_{i>j}^N (2 J^{z}_{i,j} \sigma^z_i \sigma^z_j
                    - J^{xy}_{i,j} \sigma^x_i \sigma^x_j
                    - J^{xy}_{i,j} \sigma^y_i \sigma^y_j ) ,
\end{eqnarray}
where
\begin{eqnarray} \label{Jz}
   J_{i,j}^{z} &=& \left( \frac{g}{2} \right)^2 \xi_{i,j} \varepsilon^2
               = \left( \frac{g}{2} \right)^2
                 \frac{\hbar e^4 b^2}{16 \pi \epsilon_0 m_e^4
                       \omega_z^4 d_{i,j}^3}, \\
   \label {Jxy_bis}
   J_{i,j}^{xy} &=&  \left(\frac{g}{4}\right)^2 \xi_{i,j}
                 \varepsilon^2 \frac{\omega_z^4}{\omega_a^2 \tilde{\omega}_c^2}
                 \simeq 10^6 \left(\frac{g}{4}\right)^2 \frac{%
                   \hbar e^4 b^2}{16 \pi \epsilon_0 m_e^4 \omega_c^4 d_{i,j}^3}.
\end{eqnarray}
In Eq.~(\ref{Jxy_bis}) we used the
relations $\omega_a \simeq 10^{-3} \omega_c$ and
$\tilde{\omega}_c\simeq\omega_c$.
We obtain a spin-spin interaction that is antiferromagnetic (ferromagnetic)
if it is transmitted by the cyclotron (axial) motion.

The situation is completely different when
the linear array of electrons is aligned along the $x$ axis
\begin{eqnarray}\label{Heff2}
H'_s &\simeq& \sum_{i=1}^{N}\frac{\hbar}{2}\omega_{s} \sigma^z_i +
\frac{\hbar}{2} \sum_{i<j}^N
        \left( 2 J^{z}_{i,j} \sigma^z_i \sigma^z_j
             - J^{xy}_{i,j} \sigma^x_i \sigma^x_j
             - J^{xy}_{i,j} \sigma^y_i \sigma^y_j
        \right).
\end{eqnarray}
In this case, the sign of the Heisenberg like coupling is reversed.
The ferromagnetic (antiferromagnetic) interaction is associated
to the cyclotron (axial) motion.
Similar results were also found in the case of ions, in a Paul
trap,  driven
by six counterpropagating laser beams \cite{porras}.
\section{A channel for quantum communication}
\label{chain}
The Hamiltonians (\ref{Heff1}) and (\ref{Heff2}) describe a system
of $N$ spins coupled through Heisenberg-like interactions. These
Hamiltonians can transmit an unknown spin state, from the
electron placed at one end of the linear array, to the electron
placed at the other end of the array. The remarkable fact is that
this quantum information transfer is realized only by means of the
free dynamical evolution of the system, without requiring any
external action by the experimenter during the transfer.

Therefore, let us analyze the potentialities of our system as a quantum
communication channel.
In our scheme, the dependence of the spin-spin coupling
strength on the system parameters is shown in Eqs.~(\ref{Jz})
and (\ref{Jxy_bis}).
In particular, $J_{i,j}^{z}$, $J_{i,j}^{xy}$ are proportional to
$1/d_{i,j}^3$, that is they decrease with the distance between the
particles $i$ and $j$ according to the dipolar decay law. They
also depend on the applied magnetic field gradient and on the
characteristic frequencies of the electron motion.
More specifically,
the value of $J^{xy}_{i,j}$ ($J^{z}_{i,j}$) depends on the
cyclotron (axial) frequency $\omega_c$ ($\omega_z$). As a
consequence of this fact we can neglect $J^{xy}_{i,j}$ with
respect to $J^{z}_{i,j}$ when the value of the ratio
$\omega_c/\omega_z$ is sufficiently large, as in the case
considered in \cite{nuovo}.
Differently, in this paper, we choose
smaller values for the ratio $\omega_c/\omega_z$ (generally about
$20$ or less), so that $J^{xy}_{i,j}$ is of the same order of
magnitude of $J^{z}_{i,j}$ or even larger.
Indeed, one can easily check, from Eqs.~(\ref{Jz}) and (\ref{Jxy_bis}),
that when $\omega_c/\omega_z \simeq$~$18.8$ it is possible to obtain an
isotropic Heisenberg-like interaction with $2J^z_{i,j}=J^{xy}_{i,j}$.

Generally the time required to transfer a qubit, encoded in the spin state,
along
a Heisenberg chain depends on the values of $J_{i,j}^{xy}$, so
that the larger the value of $J_{i,j}^{xy}$ the faster the
transfer. Indeed, the state transfer time $t_{ex}$ in a Heisenberg
chain, consisting of just two spins, is equal to
\begin{equation}
t_{ex}\equiv \frac{\pi}{4J^{xy}}.
\end{equation}
We assume that the particles in our linear array are equally
spaced with $d \equiv d_{i,i+1}$ and $J^{xy} \equiv
J^{xy}_{i,i+1}$. From Eq. (\ref{Jxy_bis}) we see that $J^{xy}
\propto b^2/(\omega_c^4 d^3)$. Hence, to speed up the transfer
process we have to miniaturize the system, to increase the
strength of the magnetic field gradient and to reduce the
cyclotron frequency. However, the value of the cyclotron frequency
$\omega_c$, depending on the confining magnetic field, cannot be
decreased at will, since it should be sufficiently large to cool
the cyclotron motion to its ground state. For example, at the trap
temperature of $80$~mK \cite{gabrielse} it is sufficient a
cyclotron frequency of the order of $10$~GHz. The inter-particle
distance $d$ depends on the level of miniaturization of the trap.
We consider $d$ varying from few microns to 50~$\mu$m. Finally
stronger local magnetic gradients are, in
general, achievable by reducing the micro-trap size.

Essentially, the effective Heisenberg-like Hamiltonians,
Eqs.~(\ref{Heff1}) and (\ref{Heff2}), have been
obtained by taking two steps: we applied an appropriate unitary
transformation and then disregarded the residual coupling between
the different degrees of freedom. Both these steps, in general,
introduce errors which reduce the accuracy of our system in
reproducing an array of particles interacting according to the
Heisenberg model.
In particular, we neglected terms representing a
residual coupling between the spin and the motional degrees of
freedom, as well as between the different spatial oscillators.
In Appendix~\ref{appendix}, we analyse in detail the role of each of
these terms.
Here, we only present the most relevant part of this interaction
\begin{equation} \label{res_coup}
   H_r \simeq \varepsilon^2 \sum_{i=1}^N \hbar\omega_z
                \left[ \frac{\omega_z^2}{4\omega_c \omega_a}
                      -\frac{\omega_z^2}{4 \omega_c^2} a_{m,i}^\dagger a_{m,i}
                      +\left( \frac{\omega_z^2}{4 \omega_c\omega_a}-1
                       \right) a_{c,i}^\dagger a_{c,i}
                \right] \sigma_i^z ,
\end{equation}
which
affects the spin frequency, introducing a dependence on the cyclotron and
magnetron motion.
As a consequence each particle acquires a different
spin frequency with a finite linewidth due to the thermal state of the
motional degrees of freedom.

In order to know how precisely our model can simulate an
ideal Heisenberg system we introduce the system fidelity
\begin{equation}\label{fidelity}
   \mathcal{F} \equiv \langle\psi_f|\mbox{Tr}_{ext}[\rho(t)]|\psi_f\rangle ,
\end{equation}
where
\begin{equation} \label{psi_sf}
   |\psi_f \rangle \equiv e^{-\frac{i}{\hbar}H_s
t}|\psi_0\rangle,
\end{equation}
with $|\psi_0\rangle$ being the initial state of the spin chain
and $H_s$
is the Heisenberg Hamiltonian Eq.~(\ref{Heff1}).
The operator $\rho(t)$ in Eq.~(\ref{fidelity}) represents the
density operator of the electron chain, including the motional degrees
of freedom, evolved at the time $t$
according to the full Hamiltonian of the system Eq.~(\ref{Harray}).
We also
assume that initially the axial, cyclotron, and magnetron motions
are prepared in thermal mixtures with, respectively, an average excitation
number
$\bar{k}$, $\bar{n}$ and $\bar{l}$.
The reduced density operator, describing the spin state,
is then obtained by tracing over
the spatial modes of the electrons.

The system fidelity
can be analytically calculated.
The details are provided in Appendix~\ref{appendix}.
In general, the fidelity can be written as
\begin{equation}
 \mathcal{F} = 1 - \mathcal{E}_r - \varepsilon^2 \mathcal{E}_S,
\end{equation}
where $\mathcal{E}_r$ and $\mathcal{E}_S$ represent, respectively,
the errors due to the residual coupling, Eq.~(\ref{res_coup}),
and to the canonical transformation.
In the simplest case of just two electrons, we find
\begin{equation}  \label{fidres}
  \mathcal{E}_r =1 -
   \sum_{n_1,l_1}\sum_{n_2,l_2}P_{\bar{n}}(n_1)P_{\bar{l}}(l_1)
   P_{\bar{n}}(n_2)P_{\bar{l}}(l_2)
\left[\mathcal{F}_d\left(\frac{\delta_s(n_1,l_1,n_2,l_2)}{4 J^{xy}}\right)
\right] ,
\end{equation}
with $P_{\overline{m}}(m)$, Eq.~(\ref{Pm}), being the
occupation probability for the $m$th Fock state,
\begin{equation}
  \mathcal{F}_d(\zeta) = \frac{1}{3}
     \left[ 1 + \frac{\cos(\frac{\zeta \pi}{2})
            \sin(\frac{\pi}{2} \sqrt{1+\zeta^2})}{\sqrt{1+\zeta^2}}
            + \frac{\sin^2(\frac{\pi}{2} \sqrt{1+\zeta^2})}{1+\zeta^2}
     \right]
\end{equation}
and
\begin{equation}
  \delta_s \equiv \varepsilon^2 \omega_z
      \left[
        \left(\frac{\omega_z^2}{2 \omega_c \omega_a}-2 \right)(n_2-n_1)-
        \frac{\omega_z^2}{2 \omega_c^2}(l_2-l_1)
      \right]
\end{equation}
being the detuning between the two spin frequencies.
The fidelity decreases because of this finite detuning, which is determined
by the thermal state of the cyclotron and magnetron oscillators.
Indeed, the error, Eq.~(\ref{fidres}), vanishes in the ideal case of
zero detuning $\delta_s =0$. \\
Also the error due to the canonical transformation
\begin{equation}\label{fid}
 \mathcal{E}_S = \frac{1}{3} (2\bar{k}+1)
               + \frac{\omega_z^3}{6\omega_a^2 \omega_c} (5 \bar{n}+1)
               + \frac{\omega_z^3}{6(\omega_s-\omega_m)^2\omega_c} (5
                  \bar{l}+4)
\end{equation}
becomes larger when the electron motion is relatively \emph{hot}.
From Eq.~(\ref{fid}), we see that this error is proportional to the
average excitation numbers $\bar{k}$, $\bar{n}$, and $\bar{l}$.
\begin{table} \label{tab}
{\begin{tabular}{  c |  c   c   c   c   |  c   c   c} \hline
$ $ & $ $ & $A$ & $ $ & $ $  &$B$   & $ $ & $ $\\
\hline
$d$ ($\mu$m)           &$50$      &$30$       & $10$     & $3$           &$10$   &$3$\\
$\omega_z/2\pi$(MHz)   &$490$     &$490$      & $490$    & $1200$        &$730$  &$4500$\\
$b$(T/m)               &$350$     &$600$      & $1800$   & $1800$        &$1100$  &$1100$\\
$\bar{l}$              &$0.01$    &$0.1$      & $2$      & $50$          &$0.15$  &$1$  \\
$J^{xy}$ (kHz)         &$0.01$    &$0.14$     & $35$     & $1300$        &$2.5$   &$100$\\
\hline
\end{tabular}}
\caption{\label{table} Table showing the values of the axial
frequency $\omega_z/2\pi$, the magnetic gradient $b$, the average
magnetron excitation number $\bar{l}$ and the coupling strength
$J^{xy}$ for different choices of the nearest neighbor distance
$d$. In case A (B) we have $\mathcal{F}=0.99$ ($\mathcal{F}=0.999$) and
$\omega_c/2\pi=8$~GHz ($\omega_c/2 \pi=11$~GHz). We suppose that the
axial and cyclotron motion are thermalized with the trap
environment at the temperature of $80$~mK.}
\end{table}
Therefore, to increase the system fidelity it is essential to cool,
possibly to the ground state, the electron motion.
This comes automatically for the cyclotron oscillator, when the
trap is at a temperature below 1~K, whereas the cooling of the
axial and magnetron oscillators requires appropriate techniques
\cite{brown,powell}.

We present a number of cases in Table \ref{table}, when the
fidelity approaches the value one.
We see that, for $\mathcal{F}=0.99$
($\mathcal{F}=0.999$) and the inter-particle distance
$d$ ranging from 50~$\mu$m (10~$\mu$m) to few
microns, we have a coupling constant
$J^{xy}$ in the range 10~Hz $\div$ 1.3~MHz
(2.5~kHz $\div$ 100~kHz).
For example, in the case of $d=10$~$\mu$m
we obtain $J^{xy}=35$~kHz by taking a cyclotron frequency
$\omega_c/2 \pi=8$~GHz,
an axial frequency $\omega_z/2 \pi=490$~MHz,
and a magnetic gradient $b=1800$~T/m.

We
also recall that the decoherence time of the spin state as well as
the heating time of the spatial motions, estimated according to the
model described in \cite{ciaramicoli_5,turchette,henkel}, is much
longer than the transfer time $t_{ex}$.
This remains true also for moderate values of the coupling strength
$J^{xy}$, thus allowing the transmission of the qubit state across
the chain within
the decoherence time of the system.

Finally we note that our system offers the possibility, in
principle, to apply arbitrary site specific changes to its
parameters, such as the inter-particle distance, the magnetic
gradient strength, and the magnetic field magnitude. Hence, as
suggested in \cite{kay,avellino}, by means of these local
modifications one can optimize the transmission rate and the
fidelity of the chain.
\section{Conclusions}
\label{concl}
In this paper we presented a scheme for implementing a spin chain
with long range interactions by means of a linear array of
electrons confined in micro-Penning traps. Both antiferromagnetic
and ferromagnetic Heisenberg-like systems can be realized using a
local magnetic field gradient, mediated by the electrostatic interaction
between the trapped particles. In particular, we derived an
analytical formula for the strength of the spin-spin coupling,
which determines the transmission rate of the channel, as a
function of the relevant system parameters like the inter-particle
distance, the cyclotron frequency, and the value of the applied
magnetic gradient. In our analysis we also estimated the fidelity
of the system in reproducing an effective Heisenberg chain, by
taking into account the effects produced by the coupling
between the different degrees of freedom of the particles. We
found that the fidelity depends on the frequency and the amplitude
of the spatial motion of the particles. In general, higher values
of the fidelity are obtained for smaller values of the spatial
motion amplitudes and for larger values of the detuning between the
characteristic trapping frequencies.
The numerical estimates, calculated for an inter-particle
distance $d$ varying from
50~$\mu$m to few microns, give a spin-spin coupling strength
$J^{xy}$ in the
range 10~Hz $\div$ 1.3~MHz with a fidelity of $99\%$.
Even in the case of a relatively weak coupling constant, the transmission of the
qubit state from one end to the other of the chain takes place well
within the decoherence time of the system.
Moreover, the geometry of the
system offers the possibility to apply arbitrary site-specific
changes of its parameters in order to optimize the transmission
rate and the fidelity of the quantum channel.

In conclusion, an array of electrons confined in micro-Penning traps
lends itself to implement, within
the reach of current technology, quantum channels with high
accuracy and sufficiently large transmission rates.
Furthermore, the versatility of our scheme allows one to simulate also more
general spin systems, in one and two dimensions, thus paving the way
towards the observation of quantum phase transitions.
\appendix
\section{Fidelity}
\label{appendix}
In this appendix we provide a brief description of the approach
adopted to estimate the fidelity, as defined in
Eq.~(\ref{fidelity}).
Our starting point is the complete single
electron Hamiltonian Eq.~(\ref{Hglo3}).
In order to remove from this
Hamiltonian, to the first order in $\varepsilon$, the coupling
between the different particle motions, we should apply a unitary
transformation which takes into account also the magnetron oscillator
\begin{eqnarray} \label{S2}
S &=& \varepsilon \sum_{i=1}^N
      \left[
             \left( \frac{g}{4} \sigma^z_{i} + a_{c,i}^\dagger a_{c,i}+
                    \frac{\omega_m}{\omega_c}a_{m,i}^\dagger a_{m,i}
             \right) (a_{z,i}^\dagger-a_{z,i})
             + \frac{g}{4} \frac{\omega_z}{\omega_{a}}
             \sqrt{\frac{\omega_z}{\tilde{\omega}_{c}}}
             \left( \sigma^{(-)}_{i} a_{c,i}^\dagger -\sigma^{(+)}_{i} a_{c,i}
             \right)
      \right. \nonumber \\
&+&   \left. \frac{g}{4} \frac{\omega_z}{\omega_{s}-\omega_m}
             \sqrt{\frac{\omega_z}{\tilde{\omega}_{c}}}
             \left( \sigma^{(-)}_{i} a_{m,i}-\sigma^{(+)}_{i}
                    a_{m,i}^\dagger
             \right)
      \right].
\end{eqnarray}
This unitary transformation represents a
generalization of the transformation Eq.~(\ref{S}), since it encompasses
all the degrees of freedom of the particles.\\
From the definition of the fidelity, Eq.~(\ref{fidelity}), we
can write \cite{porras}
\begin{equation}\label{fide2}
\mathcal{F} = \langle\psi_{f}|\mbox{Tr}_{ext}[e^{-S} e^{-\frac{i}{\hbar} H' t}
            e^S \rho(0) e^{-S}
            e^{\frac{i}{\hbar} H' t} e^S] |\psi_{f}\rangle ,
\end{equation}
where
\begin{equation}
   H' \equiv H_{ext} + H_s + H_r ,
\end{equation}
with
\begin{equation}
H_{ext} = \sum_{i=1}^{N}
          \left(  -\hbar \omega_m a^\dagger_{m,i} a_{m,i}
                  +\hbar \omega_c a^\dagger_{c,i} a_{c,i}
                  +\hbar \omega_z a^\dagger_{z,i} a_{z,i}
          \right)
\end{equation}
being the Hamiltonian describing the uncoupled external dynamics of the
particles.
The spin Hamiltonian $H_s$ is given in Eq. (\ref{Heff1}), whereas
 $H_r$ includes the residual
coupling between the spin and the spatial degrees of freedom
\begin{eqnarray} \label{Hr}
   H_r &\simeq& \varepsilon^2 \sum_{i=1}^N \hbar\omega_z
        \left\{
                \left[ \frac{\omega_z^2}{4\omega_c \omega_a}
                      -\frac{\omega_z^2}{4 \omega_c^2} a_{m,i}^\dagger a_{m,i}
                      +\left( \frac{\omega_z^2}{4 \omega_c\omega_a}-1
                       \right) a_{c,i}^\dagger a_{c,i}
                \right] \sigma_i^z
        \right. \nonumber \\
    &-& \frac{1}{2}  \sqrt{\frac{\omega_z}{\omega_c}}
        \left( a_{z,i}^\dagger-a_{z,i} \right)
        \left[ \sigma^{(+)}_i a_{c,i} -\sigma^{(-)}_i a_{c,i}^\dagger
               + \frac{3}{2} \left( \sigma^{(+)}_i a_{m,i}^\dagger
                                    - \sigma^{(-)}_i a_{m,i}
                             \right)
        \right] \nonumber \\
    &+& \left. \frac{\omega_z^2}{8 \omega_c \omega_a}
               \left( a_{c,i} a_{m,i} + a_{c,i}^\dagger a_{m,i}^\dagger
               \right) \sigma_i^z
        \right\} +\varepsilon \sum_{i\neq j}^N \hbar \xi_{i,j} \left[ g \left(a_{z,i}+a_{z,i}^\dagger\right)\sigma_j^z
        \right. \nonumber \\
    &-& \left.\frac{g}{2}
   \left(\frac{\omega_z}{\tilde{\omega}_{c}}\right)^{\frac{3}{2}} \frac{\omega_z}{\omega_a}
    \left(a_{c,i}\sigma^{(+)}_j +a_{c,i}^\dagger\sigma^{(-)}_j\right)
    \right].
\end{eqnarray}
We assume that initially the cyclotron, axial, and magnetron
oscillators are in a thermal mixture, each one represented by the usual density
operator
\begin{equation}
 \rho_{th} = \sum_{m=0}^{+\infty}
             P_{\overline{m}}(m) | m \rangle \langle m |,
\end{equation}
with \cite{schleich}
\begin{equation} \label{Pm}
P_{\overline{m}}(m) \equiv \left(\frac{1}{1+\overline{m}}\right)
                        \left(\frac{\overline{m}}{1+\overline{m}}\right)^{m}
\end{equation}
being the occupation probability of the $m$th Fock state of a
harmonic oscillator with average excitation number $\overline{m}$.
The initial spin state of the chain is
\begin{equation} \label{psi0}
   | \psi_0 \rangle \equiv \left(\cos \frac{\theta}{2} \, |\downarrow\rangle_1+
                            e^{i\phi}\sin\frac{\theta}{2} \,|\uparrow\rangle_1
                           \right)
             |\downarrow\rangle_2 \ldots |\downarrow\rangle_N .
\end{equation}
The information is stored in the state of the first qubit and should be
transmitted to the opposite end of the chain, to the $N$th spin.
Therefore, the density operator of the system at time $t=0$ is
\begin{equation}
  \rho(0) \equiv \rho^{(spin)} \otimes \rho^{(ext)} .
\end{equation}
The ideal final state of the spin chain is represented by the state vector
\begin{equation}
| \psi _f \rangle = \exp\left(-\frac{i}{\hbar} H_s t \right) | \psi_0 \rangle,
\end{equation}
which is obtained from the initial spin state, Eq.~(\ref{psi0}), when the
system is described by the Heisenberg Hamiltonian $H_s$, Eq.~(\ref{Heff1}).

Now to calculate the value of the fidelity, we make an expansion of
Eq.~(\ref{fide2}) in
powers of $S$ and consider terms up to the second order in
$\varepsilon$
\begin{eqnarray}\label{FS}
\mathcal{F} & \simeq &  \langle A\rho A^{-1} \rangle
    +  \frac{1}{2} \langle A \rho A^{-1} S^2 +A \rho S^2 A^{-1} +AS^2 \rho
                           A^{-1}+S^2A \rho A^{-1}
                   \rangle \nonumber \\
    &-& \langle A \rho SA^{-1}S - AS \rho A^{-1}S + AS\rho SA^{-1}
                  + SA\rho A^{-1} S - SA\rho SA^{-1} + SAS\rho A^{-1}
        \rangle ,
\end{eqnarray}
where we defined $A\equiv \exp[-(i/\hbar) H' t]$,
$\rho\equiv \rho(0)$ and $\langle\ldots\rangle\equiv\langle
\psi_{f}| \mbox{Tr}_{ext}[\ldots]|\psi_{f}\rangle$.
The first order terms in $S$ have been
omitted since their contribution, after tracing over the spatial
degrees of freedom, is zero.

In the absence of the residual couplings, contained in the Hamiltonian
$H_r$, the spin chain evolution is unaffected by the thermal state of
the motional degrees of freedom, because $[H_{ext},H_s]=0$.
This leads to $\langle A \rho A^{-1} \rangle=1$.
The corrections to the fidelity come both from the presence of $H_r$ and
from the canonical transformation, represented by the
remaining ten terms of Eq.~(\ref{FS}).
In order to separate the two effects, we first evaluate the impact of the
unitary transformation when $H' \simeq H_{ext} + H_s$.
This greatly simplifies the procedure and allows to achieve an analytical
expression for the fidelity
\begin{equation} \label{fide_err}
\mathcal{F} \simeq \langle A\rho A^{-1} \rangle
                   - \varepsilon^2 \mathcal{E}_S,
\end{equation}
where
\begin{eqnarray}
\mathcal{E}_S &\simeq & \sum_{i=1}^N
   \left\{
       \left[ \left(\frac{g}{4}\right)^2
              \left(2 - |\langle \sigma_i^z \rangle_0|^2
                      - |\langle \sigma_i^z \rangle_f|^2
              \right) + \frac{g}{2} \left(\langle \sigma_i^z \rangle_0
              - \langle \sigma_i^z \rangle_f \right)
              \left( \bar{n} + \frac{\omega_m}{\omega_c} \bar{l} \right)
       \right] (2\bar{k}+1)
   \right. \nonumber \\
   &+& \left(\frac{g}{4}\right)^2 \frac{\omega_z}{\tilde{\omega}_c}
       \left[\frac{\omega_z^2}{\omega_a^2} \, (2\bar{n}+1 +
             \langle \sigma_i^z \rangle_0) + \frac{\omega_z^2}{(\omega_s-
             \omega_m)^2} \, (2\bar{l}+1 - \langle \sigma_i^z \rangle_0)
       \right. \nonumber \\
&-&\left.
       \left.
          \left(\frac{\omega_z^2}{\omega_a^2} \, (2\bar{n}+1) +
                \frac{\omega_z^2}{(\omega_s-\omega_m)^2} \, (2\bar{l}+1)
          \right)
          \left(\langle \sigma_i^{(-)} \rangle_0
                \langle \sigma_i^{(+)} \rangle_0
               +\langle \sigma_i^{(-)} \rangle_f
                \langle \sigma_i^{(+)} \rangle_f
          \right)
       \right]
    \right\}.
\end{eqnarray}
The expectation values $\langle \ldots \rangle_0$ and
$\langle \ldots \rangle_f$ are calculated, respectively, over the initial
and final state of the spin chain.
At the swapping time, when the state of the first spin has moved to the other
end of the chain,
\begin{eqnarray}
   \sum_{i=1}^N \langle \sigma_i^z \rangle_0 &=&
   \sum_{i=1}^N \langle \sigma_i^z \rangle_f = - (N-1) - \cos\theta, \\
   \sum_{i=1}^N |\langle \sigma_i^z \rangle_0|^2 &=&
   \sum_{i=1}^N |\langle \sigma_i^z \rangle_f|^2 = N-1 + \cos^2\theta, \\
   \sum_{i=1}^N \langle \sigma_i^{(-)} \rangle_0 &=&
   \sum_{i=1}^N \langle \sigma_i^{(-)} \rangle_f =
                \frac{e^{i\phi}}{2} \sin\theta , \\
   \sum_{i=1}^N \langle \sigma_i^{(+)} \rangle_0 &=&
   \sum_{i=1}^N \langle \sigma_i^{(+)} \rangle_f =
                \frac{e^{-i\phi}}{2} \sin\theta . \\
\end{eqnarray}
Moreover, after
averaging over all the initial states in the Bloch sphere,
i.e. evaluating
$(1/4 \pi) \int_0^{\pi} \int_0^{2 \pi} \mathcal{E}_S \, \sin\theta \,
d\theta \, d\phi$, we obtain
\begin{equation}\label{fide3}
\mathcal{E}_S \simeq \frac{1}{3} (2\bar{k}+1) +
                     \frac{1}{6}\frac{\omega_z}{\omega_c}
  \left[ \frac{\omega_z^2}{\omega_a^2} \left(2\bar{n}+1 + 3(N-1)\bar{n}\right)
        +\frac{\omega_z^2}{(\omega_s-\omega_m)^2}
               \left(2\bar{l}+1 + 3 (N-1) (\bar{l}+1) \right)
  \right],
\end{equation}
where $\bar{k}$, $\bar{n}$, and $\bar{l}$ denote, respectively, the
average axial, cyclotron and magnetron excitation number.
The expression (\ref{fide3}) gives the error due to
the unitary transformation.

Let us consider now the effects of the Hamiltonian $H_r$,
contained in the term $\langle A \rho A^{-1} \rangle$ of
Eq.~(\ref{fide_err}).
The residual couplings produce mainly two effects: they induce
transitions between the motional states of the electron and make
the electron spin frequency depend on the state of the external degrees
of freedom.
Both these effects, as we will see, reduce the system
fidelity.

The probability to observe transitions between the states of the
electron motion can be easily estimated using a perturbative
approach. Indeed, the probability for the transition
$|\psi_m\rangle \rightarrow |\psi_n\rangle$ is not larger than
roughly $4|\langle \psi_n |\Delta H|\psi_m \rangle|^2/(\hbar
\omega_{nm})^2$, where $\Delta H$ is the perturbation,
$\omega_{nm}$ the transition frequency and $|\psi_i\rangle$ the
$i$-th eigenstate of the unperturbed Hamiltonian. In our case, the
Hamiltonian $H_r$ plays the role of $\Delta H$ and the terms,
responsible for the transitions between electronic states, are in
the last three lines of Eq.~(\ref{Hr}). For example, the terms
proportional to $a_{z,i} \sigma^{(+)}_i a_{c,i}$ induce
transitions between the eigenstates $|n,k,l,\downarrow \rangle$
and $|n - 1,k - 1,l,\uparrow\rangle$ of the single electron
Hamiltonian
\begin{equation}\label{seH}
H_{0}=-\hbar \omega_m a_m^\dagger a_m+\hbar \omega_c a_c^\dagger
a_c+\hbar \omega_z a_z^\dagger a_z+\frac{\hbar}{2}\omega_s
\sigma^z
\end{equation}
with probability of the order of
$\varepsilon^4[\omega_z^3/\omega_c(\omega_z-\omega_a)^2]
 k n$.
A similar perturbative approach allows us to estimate also the
transition probability due to the other terms of Eq. (\ref{Hr}).
These probabilities, for the terms involving couplings between
different motions of the same particle, are proportional to
$\varepsilon^4/(\Delta \omega)^2$ where $\Delta \omega$ denotes
the detuning between the electron oscillation frequencies. Hence,
the error, in this case, is always negligible because is a
correction of the fourth order in $\varepsilon$ and, moreover, the
characteristic frequencies of the electron motion are quite
different from each other. Very small errors are also produced by
the terms in Eq. (\ref{Hr}) involving the dynamics of different
particles. In this case the transition probabilities are of the
order of $\varepsilon^2 (\xi_{i,j}/\omega_z)^2$ and $\varepsilon^2
(\xi_{i,j}/\omega_z)^2(\omega_z/\omega_c)^3(\omega_z/\omega_a)^4$.
Indeed, these values, for our choices of the system parameters,
are negligible.

In addition to the state transitions, the residual couplings
enforce a dependence of the spin frequency on the state of the
particle motion. The correction $\Delta \omega_s$ to the spin
frequency
\begin{eqnarray} \label{delta_omega_s}
\Delta \omega_s (n,l) \simeq \varepsilon^2 \omega_z
       \left[
              \frac{\omega_z^2}{2 \omega_c \omega_a} +
              \left(\frac{\omega_z^2}{2 \omega_c \omega_a}-2 \right)n -
               \frac{\omega_z^2}{2 \omega_c^2} l
       \right]
\end{eqnarray}
depends on the cyclotron and
magnetron excitations.
Indeed, the constant shift proportional to $\omega_z^2 / 2 \omega_c \omega_a$
equally affects all the spins in the chain and, therefore, does not introduce
any detuning between the spin frequencies.
This term can be taken into account by redefining the spin precession
frequency $\omega_s$.
On the contrary, the last two terms of
Eq.~(\ref{delta_omega_s}) introduce a detuning between the spin
frequencies along the chain, since
the cyclotron and magnetron
oscillators are in a thermal mixture with fluctuating excitation numbers
$n$ and $l$.
As a consequence each spin in the chain acquires a different frequency
depending on
the thermal state of the electron motion.
This leads, as we will show, to a reduction of the system fidelity.

For the sake of simplicity, we restrict our analysis to the case of just
two electrons in the chain.
The corresponding Hamiltonian reads
\begin{equation}  \label{Hsd}
H_{sd} = \frac{\hbar}{2} \omega_1 \sigma^z_1 +
         \frac{\hbar}{2} \omega_2 \sigma^z_2
         + 2 \hbar J^{xy} \left( \sigma^{(+)}_1 \sigma^{(-)}_2
                                +\sigma^{(-)}_1 \sigma^{(+)}_2
                          \right)
         - 2 \hbar J^{z} \sigma^z_1 \sigma^z_2,
\end{equation}
with $\omega_i = \omega_s +\Delta\omega(n_i,l_i)$.
The unitary evolution of the system gives at the swapping time
$t_{ex} = \pi / 4 J^{xy} $
\begin{eqnarray} \label{Hd1}
|\downarrow\rangle_1 |\downarrow\rangle_2 &\rightarrow&
e^{2 i J^z t_{ex}} e^{\frac{i}{2} (\omega_1+\omega_2) t_{ex}}
|\downarrow\rangle_1 |\downarrow\rangle_2, \\
|\uparrow\rangle_1 |\downarrow\rangle_2 &\rightarrow& e^{-2 i J^z t_{ex}}
      \left\{ - \frac{i}{\sqrt{1+\zeta^2}}
             \sin\left(\frac{\pi}{2}\sqrt{1+\zeta^2}\right)
             |\downarrow\rangle_1 |\uparrow\rangle_2
      \right. \nonumber \\
&+& \left.
          \left[ \cos\left(\frac{\pi}{2}\sqrt{1+\zeta^2}\right)+
                 \frac{i \zeta}{\sqrt{1+\zeta^2}} \sin
                 \left(\frac{\pi}{2}\sqrt{1+\zeta^2}\right)
          \right]|\uparrow\rangle_1 |\downarrow\rangle_2
    \right\},  \label{Hd2}
\end{eqnarray}
where $\zeta\equiv\delta_s/(4 J^{xy})$ with
\begin{equation}
\delta_s(n_1, l_1, n_2, l_2)
         \equiv  \omega_2 - \omega_1
         = \varepsilon^2 \omega_z
               \left[ \left( \frac{\omega_z^2}{2\omega_c\omega_a}-2 \right)
                      (n_2 - n_1) - \frac{\omega_z^2}{2\omega_c^2}(l_2 - l_1)
               \right] .
\end{equation}
Hence, by using the relations (\ref{Hd1})
and (\ref{Hd2}), we obtain the system fidelity
\begin{equation}
\mathcal{F}_d(\zeta) = \frac{1}{3}
    \left[ 1 + \frac{\cos(\frac{\zeta \pi}{2})
           \sin(\frac{\pi}{2} \sqrt{1+\zeta^2})}{\sqrt{1+\zeta^2}}
         + \frac{\sin^2 (\frac{\pi}{2} \sqrt{1+\zeta^2})}{1+\zeta^2}
    \right]
\end{equation}
when the cyclotron and magnetron oscillators are in the
states $|n_i, l_i\rangle$, with $i = 1, 2$.
Consequently, in the case of a
thermal mixture,
the expression for the fidelity becomes
\begin{equation} \label{fidelity_detuning}
\mathcal{F} \equiv \sum_{n_1,l_1}\sum_{n_2,l_2}
         P_{\bar{n}}(n_1)P_{\bar{l}}(l_1)P_{\bar{n}}(n_2)P_{\bar{l}}(l_2)
         \left[\mathcal{F}_d
               \left( \frac{\delta_s(n_1,l_1,n_2,l_2)}{4 J^{xy}}
               \right)
         \right].
\end{equation}
We use this formula together with Eq.~(\ref{fide3})
to numerically evaluate the fidelity of our system.
%
\begin{acknowledgments}
This research was supported by the European Commission through the
Specific Targeted Research Project \emph{QUELE}, the Integrated Project
FET/QIPC \emph{SCALA}, and the Research Training Network \emph{CONQUEST}.
\end{acknowledgments}
%

\begin{thebibliography}{99}
\bibitem{bose}S. Bose, Phys. Rev. Lett. \textbf{91}, 207901 (2003).
\bibitem{osborne}T.J. Osborne, N. Linden, Phys. Rev. A \textbf{69}, 052315 (2004).
\bibitem{chri1}M. Christandl, N. Datta, A. Ekert, and A. J. Landahl, Phys. Rev. Lett. \textbf{92}, 187902 (2004).
\bibitem{chri2}M. Christandl, N. Datta, T.~C. Dorlas, A. Ekert, A. Kay,
and A.~J.
Landahl, Phys. Rev. A \textbf{71}, 032312 (2005).
\bibitem{burgarth1} D.Burgarth and S. Bose, Phys. Rev. A \textbf{71}, 052315 (2005).
\bibitem{burgarth2}D.Burgarth and S. Bose, New J. Phys. \textbf{7}, 135 (2005).
\bibitem{kay} A. Kay, Phys. Rev. A \textbf{73}, 032306 (2006).
\bibitem{avellino} M. Avellino, A. J. Fisher and S. Bose,
         Phys. Rev. A \textbf{74}, 012321  (2006).
\bibitem{ciaramicoli4} G. Ciaramicoli, I. Marzoli, and P. Tombesi,
         Phys. Rev. Lett. \textbf{91}, 017901 (2003).
\bibitem{ciaramicoli_5} G. Ciaramicoli, I. Marzoli, and P. Tombesi,
         Phys. Rev. A \textbf{70}, 032301 (2004).
\bibitem{nuovo} G. Ciaramicoli, F. Galve, I. Marzoli, and P. Tombesi,
         Phys. Rev. A \textbf{72}, 042323 (2005).
\bibitem{wiretrap} J. R. Castrej\'{o}n-Pita and R. C. Thompson
         Phys. Rev. A \textbf{72}, 013405 (2005).
\bibitem{stahl} S. Stahl, F. Galve, J. Alonso, S. Djekic, W. Quint,
         T. Valenzuela, J. Verd\`u, M. Vogel, and G. Werth,
         Eur. Phys. J. D \textbf{32}, 139 (2005).
\bibitem{porras} D. Porras and J. I. Cirac, Phys. Rev. Lett. \textbf{92},
        207901 (2004).
\bibitem{drndic} M. Drndi\'{c}, C. S. Lee, and R. M. Westervelt,
        Phys. Rev. B \textbf{63}, 085321 (2001).
\bibitem{brown} L.S. Brown and G. Gabrielse, Rev. Mod. Phys. \textbf{58},
         233 (1986).
\bibitem{gabrielse} S. Peil and G. Gabrielse, Phys. Rev. Lett. \textbf{83}, 1287
        (1999).
\bibitem{powell} H.F. Powell, D.M. Segal, and R.C. Thompson, Phys.
        Rev. Lett. \textbf{89}, 093003 (2002).
\bibitem{wunderlich} F. Mintert and Ch. Wunderlich, Phys. Rev. Lett.
        \textbf{87}, 257904 (2001);
        Ch. Wunderlich, \emph{Laser Physics at the Limit}
        (Springer, Heidelberg, 2001), p. 261.
\bibitem{jackson} J.D. Jackson, \textit{Classical Electrodynamics}, 2nd
         ed., (Wiley, New York, 1975).
\bibitem{turchette} Q.A. Turchette, D. Kielpinski, B.E. King, D. Leibfried, D.M.
         Meekhof, C.J. Myatt, M.A. Rowe, C.A. Sackett, C.S. Wood, W.M.
         Itano, C. Monroe, and D.J. Wineland, Phys. Rev. A \textbf{61},
         063418 (2000).
\bibitem{henkel} C. Henkel, S. P\"{o}tting, and M. Wilkens,
         Appl. Phys. B: Lasers Opt. \textbf{69}, 379 (1999).
\bibitem{schleich} W.~P. Schleich, \textit{Quantum Optics in Phase Space},
         (Wiley-VCH, Berlin, 2001).
%
\end{thebibliography}

%
\end{document}